\documentstyle[11pt,a4,ere99]{article}
\input{epsf}
\def\laq{\raise 0.4ex\hbox{$<$}\kern -0.8em\lower 0.62
ex\hbox{$\sim$}}
\def\gaq{\raise 0.4ex\hbox{$>$}\kern -0.7em\lower 0.62
ex\hbox{$\sim$}}

\newcommand{\beq}{\begin{equation}}
\newcommand{\eeq}{\end{equation}}
\newcommand{\bea}{\begin{eqnarray}}
\newcommand{\eea}{\end{eqnarray}}
\def \ep {\epsilon}
\def \Ga {\Gamma}
\def \r {\rho}
\def \ga {\gamma}
\def \al {\alpha}
\begin{document}

\title{Seeds of Large Scale Anisotropy
in Pre-Big-Bang Cosmology\\
\emph{Proceedings of the Spanish Relativity Meeting}\\ Bilbao, 1999}

\author{Mairi Sakellariadou}

\address{
Institut des Hautes Etudes Scientifiques, 91440 Bures-sur-Yvette, France\\
DARC, Observatoire de Paris,
UPR 176 CNRS, 92195 Meudon Cedex, France\\
E-mail: {\em {\tt mairi@amorgos.unige.ch}}}

\maketitle\abstract{
Within a string cosmology context, the large scale temperature
anisotropies may arise from the contribution of seeds to
the metric fluctuations. We study the cases of electromagnetic 
and axion seeds. We find that massless or very light axions 
can lead to a flat or slightly tilted blue spectrum, that fits
current data.
}

\section{Introduction}
 
I will briefly present some results~\cite{dgsv1,dgsv2}
 on the seeds of large scale 
anisotropy in the context of string cosmology. I  work on the 
pre-big-bang scenario~\cite{pbb} (PBB), defined as a particular 
model of inflation inspired by the duality properties of string 
theory. The question which I address is whether we can reproduce 
the observed amplitude and slope of the large scale temperature 
anisotropy and of large scale density perturbations within the PBB
scenario.

First-order scalar and tensor metric perturbations lead to primordial 
spectra that grow with frequency~\cite{gg}, with a normalization 
imposed by the 
string cut-off at the shortest amplified scales. These blue spectra 
have too little power at scales relevant for the observed anisotropies 
in the cosmic microwave background radiation (CMBR). In contrast, the 
axion energy spectra where found to be logarithmically diverging, 
leading to red spectra of CMBR anisotropies which are in conflict with 
observations. These results already ruled out four dimensional PBB 
cosmology.

However, if one allows for internal contracting dimensions in addition 
to the three expanding ones, the supersymmetric partner of the dilaton (the
universal axion of string theory) can lead to a flat Harrison-Zel'dovich
(HZ) spectrum of fluctuations for an appropriate relative evolution of 
the external and the compactified internal dimensions~\cite{buon}. 
Thus, the PBB scenario  may contain 
a natural mechanism for generating large scale anisotropy via the 
seed mechanism~\cite{ruth} (i.e., 
fluctuations of one-component of the energy momentum tensor can 
feed back on the metric through Einstein's equations).

In what follows, I consider the possibility that vacuum fluctuations 
of the electromagnetic and of the axion  field may act, at second
order, as scalar seeds of large scale structure and CMBR anisotropies.
The induced perturbations are isocurvature perturbations. 
More precisely, I examine 
whether the metric perturbation spectrum triggered by these seeds can be flat 
enough to match present measurements, consistently with the COBE
normalization of the amplitude on large scales, and with the normalization
imposed by the string cut-off at the shortest amplified scales.

\section{Large scale perturbations in the presence of seeds}

I will derive a general formula for
large scale  CMBR anisotropies in models where perturbations are
triggered by seeds. I consider the case of scalar perturbations.

\subsection{Cosmological perturbation theory with seeds}

We express the Fourier components of the energy momentum tensor 
of the seeds, $T_{\mu\nu}$, in terms of four scalar 
``seed functions''~\cite{r}
$f_\rho$, $f_p$, $f_v$ and $f_\pi$:
\bea
T_{00}({\bf k},\eta) &=& 
	M^2f_{\rho}({\bf k},\eta) \; ,
     \label{3seed00} \\
     T_{j0}({\bf k},\eta)  &=& 
	-iM^2k_jf_{v}({\bf k},\eta) \; ,
     \label{3seed0i} \\
   T_{ij}({\bf k},\eta)       &=& M^2\left
[\left(f_p ({\bf k},\eta)+{k^2\over 3}f_{\pi}({\bf k},\eta)
\right)\ga_{ij} -k_ik_jf_{\pi}({\bf k},\eta)\right] \;.
 \label{3seedij}
\eea
Here $M$ denotes an arbitrary mass scale, introduced for dimensional
reasons; $\eta$ denotes conformal time and $\ga$ represents a metric 
of constant curvature.

The perturbed Einstein's equations read~\cite{r}
\bea
k^2\Phi &=& 4\pi G\rho a^2 D+ \ep\left[f_{\rho} +
3(\dot{a}/a)f_v\right]
       \; , \label{2G1} \\
 \Phi +\Psi &=&-8\pi G a^2k^{-2} p\Pi -2\ep f_{\pi} \; , \label{2G3}
\eea
where $\ep \equiv 4\pi GM^2$, $a$ is the scale factor and dot denotes
derivative with respect to $\eta$. 
$\Pi$ is the anisotropic stress 
potential, $V$ is the peculiar velocity potential, $D$ (and $D_g$ which 
I will use later) is a 
gauge-invariant density perturbation variable. $\Phi, \Psi$ are
two geometric quantities, called the Bardeen potentials.
Since large scale CMBR anisotropies are induced at  recombination 
and later, we set  $\Pi=0$. 

The large scale anisotropies of CMBR are determined by the
combination $\Psi-\Phi$:
\beq
\Psi - \Phi   \sim \max
\left\{\ep f_\pi, \ep \eta^2\left(f_\r+3{\dot a \over a}f_v
\right)\right\}~. \label{comp}
\eeq

\subsection{ The seed contribution to CMBR anisotropies}

I calculate the CMBR anisotropies and their contribution to
$\Delta T/T$ via the Sachs-Wolfe effect~\cite{sw}. The
temperature perturbation reads~\cite{r}
\beq
{\delta T({\bf n})\over T} =\left[ {1\over 4}D_g +
V_{j}n^j
+\Psi -\Phi\right](\eta_{dec},{\bf x}_{dec})
   + \int_{\eta_{dec}}^{\eta_0}(\dot{\Psi}-\dot{\Phi})(\eta,{\bf
	x}(\eta))d\eta~, \label{dT0}
\eeq
where ${\bf x}(\eta)={\bf x}_0-(\eta_0-\eta){\bf n}$ is
the unperturbed photon position at time $\eta$ for an observer at
${\bf x}_0$, $\eta_0$ is the conformal time today, 
and ${\bf x}_{dec}={\bf x}(\eta_{dec})$.

The angular power spectrum of
CMBR anisotropies is expressed in terms
of the dimensionless coefficients $C_\ell$, which appear in the
expansion of the angular  correlation function in terms of the
Legendre polynomials $P_\ell$:
\beq
\left\langle{\delta T\over T}({\bf
n}){\delta T\over T}({\bf n}') \right\rangle_{{~}_{\!\!({\bf n\cdot
n}'=\cos\vartheta)}}=
  {1\over 4\pi}\sum_\ell(2\ell+1)C_\ell P_\ell(\cos\vartheta)~.
\label{cor}
\eeq
Here the brackets denote spatial average, or expectation values if
perturbations are quantized.
To determine  the $C_{\ell}$ we Fourier-transform Eq.~(\ref{dT0}),
defining
\beq
\varphi({\bf k}) = {1\over \sqrt{V}}\int_V
	\varphi({\bf x})e^{i{\bf k\cdot x}}d^3x~,
\eeq
For the
coefficients $C_\ell$ of Eq.~(\ref{cor}) we obtain:
\begin{equation}
C_\ell = {2\over \pi} \int
{\langle|\Delta_\ell ({\bf k})|^2\rangle  \over (2\ell +1)^2} k^2 dk ~,
\end{equation}
\begin{eqnarray}
\mbox {where} ~~~~~{\Delta_\ell \over 2\ell +1} 
&=&{1\over 4}D_g({\bf k},\eta_{dec})j_\ell(k\eta_0)
- j_\ell '(k\eta_0){\bf V}({\bf k},\eta_{dec})
\nonumber \\
&& + k\int_{\eta_{dec}}^{\eta_0}(\Psi -\Phi)({\bf k},\eta')
j_\ell '\left(k\eta_0-k\eta'\right) d\eta'~,
\label{Dl}
\end{eqnarray}
and $j'_\ell$ stands for the derivative of $j_{\ell}$ with respect to
its argument. On large angular scales, $k\eta_{dec}\ll 1$, the SW
contribution dominates and we obtain~\cite{dgsv2}:
\beq
C_\ell^{SW} = {2\over\pi}\int
k ^4dk\left\langle\left[\int_{\eta_{dec}}^{\eta_0}(\Psi -\Phi)({\bf k},
\eta)j_{\ell}'\left(k\eta_0-k\eta\right) d\eta\right]^2\right\rangle  .
\label{Cell}
 \eeq
We approximate~\cite{dgsv2} the Bardeen potentials
$\Psi, \Phi$  on super-horizon scales
by a power-law spectrum:
$\langle|\Psi-\Phi|^2\rangle= C^2(k)~(k\eta)^{2\ga} ~.$
Furthermore, we consider~\cite{dgsv2} models where the seed contribution
does not grow in time on sub-horizon scales. Thus,
\beq
 \Psi-\Phi \approx \left\{\begin{array}{ll}
	C(k)(k\eta)^\ga & ~,~ k\eta\ll 1\\
	C(k) & ~,~ k\eta\gg 1~.	\end{array}  \right.
\label{252}
\eeq
We further assume~\cite{dgsv2} 
that also $C(k)$ is given by a simple power law.
Thus, we have 
\beq
C(k)=\left\{\begin{array}{ll}
	{\cal N}k^{-3/2} (k/k_1)^{\al} ~,& k\le k_1\\
	0 ~,& k> k_1 ~,\end{array} \right.
\label{253}
\eeq
where  ${\cal N}$ is a dimensionless constant, and $k_1$ denotes a
comoving  cutoff scale.
Inserting Eq.~(\ref {253}) in Eq.~(\ref{Cell}), we obtain~\cite{dgsv2} 
\beq
C_\ell^{SW} \approx {\cal N}^2{2\over \pi}\int_0^{k_1} {dk\over
k}
    \left({k\over k_1}\right)^{2\al}|I(k)|^2 	
	~, 
\eeq
\beq
\mbox{where~~~~}
 I(k)     =   \int_{(k\eta)_{dec}}^1d(k\eta)
(k\eta)^{\ga}
j'_\ell(k\eta_0-k\eta) + j_\ell(k\eta_0-1) ~.
\eeq
We compare $C_\ell^{SW}$ with the inflationary result:
$C_\ell^{SW} \propto \Ga(\ell -1/2+ n/2)/\Ga(\ell+5/2-n/2)$, 
where $n$ denotes the spectral index.
The scale-invariant spectrum, as it has been found by the DMR
experiment~\cite{smootscott}, requires
$0.9\le n\le 1.5~.$
Thus, we get~\cite{dgsv2}    
\beq
 ~~~-0.05< \ga+1+\al < 0.25 ,~~~~ \ga \le -1~, ~~~~n\simeq 3+2(\al+\ga)
\label{273}
\eeq
\beq
\mbox {and} ~~ -0.05<\al<0.25 ,~~~~~ \ga > -1~, ~~~n=1+2\al.
\label{index>-1}
\eeq

\section {Seeds from string cosmology}

In this section we compute the seed functions $f_\r, f_v, f_\pi$, and
we estimate the Bardeen potentials for electromagnetic and axion
perturbations.

\subsection {Electromagnetic seeds}

Consider  a  stochastic background obtained by amplifying the 
quantum electromagnetic fluctuations of the vacuum.
For purely magnetic seeds (the electric component of the stochastic
background is rapidly dissipated, due to the conductivity of the
cosmic plasma), on super-horizon scales we obtain 
$f_v=0$, $f_\pi\gg \eta^2 f_\r$, leading to~\cite{dgsv2}
\beq
k^3\left|\Psi- \Phi\right|^2(k,\eta)\approx  
{\cal N}^2(k\eta)^{2\ga}(k/k_1)^{2\al}~,
\eeq
\bea
\mbox {with} ~~~~ \gamma &=&\left\{\begin{array}{ll}
	 -4 &, ~~~ \mu\le 3/4\\
	2\mu-11/2 ~~~~~~  &, ~~~ 3/4\le \mu\le 3/2 ~ \end{array} \right.\\
\alpha &=&\left\{\begin{array}{ll}
	 7/2 ~~~&, ~~~ \mu\le 3/4\\
	5-2\mu  ~~~~~~ &, ~~~ 3/4\le \mu\le 3/2 ~ \end{array}\right.\\
{\cal N} &=& \left((H_1/M_p)\over 4\pi\right)^2 (k_1\eta_{eq})^2
\mbox {~~~, ~~~ in both cases~. }
\eea
($\mu<3/2$ to avoid photon overproduction.)
$H_1$ is the physical cut-off scale at which the universe
becomes immediately radiation-dominated, and $M_p$ is the 
Planck mass.

Since, in both cases $\ga+1<0$, the seeds decay
fast enough outside the horizon. However, in both cases $\ga+\al=-0.5$, 
which implies $n=2$. Such a spectrum grows too fast
with frequency to fit the COBE measurements. 
The quadrupole amplitude~\cite{banday}
$Q_{rms-PS}=\sqrt{(5/4\pi)C_2} T_0=(18\pm 2)\mu K$
leads to
$
C_2 = (1.09  \pm  0.23)\times 10^{-10}  ~. 
$ 
Thus, compatibility with the COBE normalization implies~\cite{dgsv2}
\beq
(6-\al)\log_{10} (H_1/M_p) ~\laq~ 55(\al-2) -6 + \log_{10}(\ga+1)^2 ~.
\label{57}
\eeq
This constraint  is easily satisfied by a growing seed spectrum,
$\al>2$. Even in the limiting case $\al=2$, this condition is
marginally compatible even with the maximal expected value $H_1\sim
M_s\sim 5\times 10^{17}$ GeV.

\subsection {Axionic seeds}

We consider pseudo-scalar vacuum fluctuations amplified by the time
evolution of a higher dimensional background.
We first consider massless axions.
If $\mu < 3/4$, the situation is like for electromagnetic
seeds. The induced CMBR fluctuations have the wrong spectrum, but their
amplitude is sufficiently low to avoid conflict with observations.
However, if $3/4\leq\mu\leq 3/2$, we obtain~\cite{dgsv1,dgsv2}
\beq
k^3\left|\Psi- \Phi\right|^2(k,\eta)\approx  {\cal N}^2(k\eta)^{2\ga}
(k/k_1)^{2\al}~,
~~~ 
\mbox {with} ~~~ \ga=2\mu-7/2 ~~,~~\al=-2\mu+3
~.
\eeq
For $\mu=3/2$ we obtain a Harrison-Zel'dovich spectrum with amplitude
${\cal N}\simeq(H_1/M_p)^2$.
The non-conformal coupling of the axions to the metric leads to an
additional amplification of perturbations after the
matter-radiation equality. The normalization of the axion spectrum 
to the COBE amplitude imposes the constraint~\cite{dgsv1,dgsv2}
\beq
\log_{10}{H_1\over M_p}\simeq {164-116 \mu\over 1+2\mu}
~~~~~
\mbox {with} ~~~~~ 1.4<\mu<1.5 \label{414}~,
\eeq
\beq
\mbox {implying} ~~~~~~~~~~~~ 3\times 10^{-3} ~\laq~(H_1/M_p)~\laq~2.6~.
\label{415}
\eeq
This condition is perfectly compatible with $H_1\sim M_s
\sim 5\times 10^{17}$ GeV.

Let us now turn to the case of massive axions. In this case, the $f_\pi$ 
contribution to $\Phi, \Psi$ is negligible when the super-horizon
modes are already non-relativistic at the time of decoupling, and
we obtain~\cite{dgsv2} constant Bardeen potentials with
\beq
\ga=0 ~~~,~~~ \al=3-2\mu ~~~,~~~ {\cal N}=(H_1/M_p)^2(m/H_{eq})^{1/2}~,
\eeq
where $m$ denotes the axion mass. For $\mu=3/2$ we obtain a flat
Harrison-Zel'dovich spectrum.
The amplitude of perturbations is enhanced by the factor 
$(m/H_{eq})^{1/2}$. Thus, the axion mass, $m$, has to 
be bounded to avoid conflicting with the COBE normalization 
$C_2\approx 10^{-10}$.
In addition, we impose $1.4<\mu<1.5$,\ $m>H_{dec}\sim H_{eq}$, 
and we require that the present axion energy density is 
constrained by the critical energy density. 
We find that for a typical scale 
$H_1\sim M_s \sim (10^{-1}-10^{-2})M_p$, 
the maximal allowed axion mass window is~\cite{dgsv2}:
\beq
10^{-27} \;{\rm eV} ~\laq~ m ~\laq~ 10^{-17} \;{\rm eV} .
\label{515}
\eeq

\section{Conclusions}

In this talk, I briefly discussed, in the 
context of the PBB scenario, the possibility that 
the large scale temperature anisotropies may
arise from the contribution of seeds to the metric fluctuations.
In particular, I considered the cases in which the seed inhomogeneity 
spectrum is due to vacuum fluctuations of the 
electromagnetic field and of the (Kalb-Ramond) axion field .

In the first case, I showed  that ectromagnetic fluctuations lead 
to a spectrum that grows too fast with frequency to be compatible with 
COBE observations. Since the contribution of electromagnetic seeds to 
the large scale anisotropy is negligible, there are no constraints 
from the COBE normalization to the production of seeds for generating 
the galactic magnetic fields, via the amplification of
electromagnetic vacuum fluctuations due to a dynamical dilaton 
background~\cite{magn}.

In the second case, I discussed how a stochastic background of massless 
axions, produced within the context of the PBB scenario, is a possible
candidate for an explanation of the large scale anisotropy measured 
by COBE satellite. 
Regarding massive axions, I showed that if the axion mass
 is such that all modes outside the horizon at decoupling are 
already non-relativistic, then a slightly tilted blue spectrum is still
compatible with the amplitude and slope measured by COBE satellite,
provided the axion mass is inside an appropriate window, in 
the ultra-light mass region.

As a next step, one has to study the predictions of this model
regarding the acoustic peaks in the CMBR anisotropy power spectrum
and the linear dark matter power spectrum and compare them with
currently available experimental and observational data. Some preliminary
results are discussed in Ref.~[11]. The authors found~\cite{11}
a strong dependence of their predictions on the overall evolution
of extra dimensions during the PBB phase. In other words, further
experimental and observational data coming from the CMBR anisotropies
and the galaxy distribution, may provide some information about
the evolution of string theory's extra dimensions. 

\vspace*{-2pt}
\section*{Acknowledgments}
It is a pleasure to thank the organizers of the 1999 Spanish Relativity Meeting
for inviting me to present this work. Many thanks also to Ruth Durrer,
Maurizio Gasperini and Gabriele Veneziano, with whom I collaborated on the
papers I presented here at Bilbao.

\end{document}